\documentclass[pra,twoside,twocolumn,10pt,floatfix,showpacs,citeautoscript,superscriptaddress]{revtex4-2}

\usepackage{amsmath}
\usepackage{amssymb}
\usepackage{booktabs}
\usepackage{braket}
\usepackage{glossaries}
\usepackage{graphicx}
\usepackage[version=4]{mhchem}
\usepackage{siunitx}
\usepackage{tabularx}
\usepackage[dvipsnames]{xcolor}
\usepackage{multirow}

\usepackage[colorlinks=true]{hyperref}
\hypersetup{
    pdftoolbar=true,        
    pdfmenubar=true,        
    pdffitwindow=false,     
    pdfstartview={FitH},    
    pdftitle={My title},    
    pdfauthor={Author},     
    pdfsubject={Subject},   
    pdfcreator={Creator},   
    pdfproducer={Producer}, 
    pdfkeywords={keyword1, key2, key3}, 
    pdfnewwindow=true,      
    colorlinks=true,        
    linkcolor=black,      
    citecolor=blue,        
    filecolor=blue,      
    urlcolor=blue        
}

\hypersetup{pdfauthor={Petter Rosander, Erik Fransson, Nicklas Österbacka, Paul Erhart, and Göran Wahnström}}
\hypersetup{pdftitle={Untangling the Raman spectrum of cubic and tetragonal BaZrO3}}

\graphicspath{{figures/}}

\setacronymstyle{long-short}
\newacronym{bzo}{BZO}{\ce{BaZrO3}}
\newacronym{cx}{vdW-DF-cx}{van-der-Waals-density functional with consistent exchange}
\newacronym{dft}{DFT}{density functional theory}
\newacronym{dse}{DSE}{dielectric susceptibility expansion}
\newacronym{fwhm}{FWHM}{full width half maximum}
\newacronym{md}{MD}{molecular dynamics}
\newacronym{nep}{NEP}{neuroevolution potential}
\newacronym{tnep}{TNEP}{tensorial neuroevolution potential}
\newacronym{pes}{PES}{potential energy surface}
\newacronym{xc}{XC}{exchange-correlation}
\newacronym{dho}{DHO}{damped harmonic oscillator}

\newcommand{\gpumd}{\textsc{gpumd}}

\DeclareSIUnit\angstrom{\text {Å}}
\renewcommand{\vec}[1]{\ensuremath\boldsymbol{#1}}
\renewcommand{\epsilon}[0]{\varepsilon}

\newcommand{\overbar}[1]{\mkern 1.5mu\overline{\mkern-1.5mu#1\mkern-1.5mu}\mkern 1.5mu}

\newcommand{\addchalmers}{
    Department of Physics,
    Chalmers University of Technology,
    SE-41296, Gothenburg, Sweden
}

\begin{document}

\title{
    Untangling the Raman spectrum of cubic and tetragonal \texorpdfstring{\ce{BaZrO3}}{BaZrO3}
}

\author{Petter Rosander}
\author{Erik Fransson}
\author{Nicklas \"Osterbacka}
\author{Paul Erhart}
\author{G\"oran Wahnstr\"om}

\affiliation{\addchalmers}

\date{\today}

\begin{abstract}
Raman spectroscopy is a widely used experimental technique to study the vibrational properties of solids.
Atomic scale simulations can be used to predict such spectra, but reliable studies at finite temperatures are challenging, mainly due to the requirement of accurate and computationally efficient models for the dielectric susceptibility.
Here, we have used \gls{md} simulations together with a \gls{dft} based model for the dielectric susceptibility to determine the Raman spectrum of barium zirconate, \gls{bzo}, a well-studied oxide perovskite.
At ambient conditions, where the system is cubic, we find excellent agreement with experimentally measured Raman spectra.
Our study establishes that the relatively sharp spectra seen experimentally are due to second-order scattering.
At higher pressures, where \gls{bzo} is tetragonal, all first-order Raman active modes are identified.
Additionally, slightly below the phase transition, in the cubic phase, a broad ``central Raman peak'' appears.
The origin of this type of peak is controversial and extensively debated in connection with the dynamics of the halide perovskites.
Here, we show that it is also present in a ``hard'' oxide perovskite, and it originates from the highly overdamped R-tilt mode in the cubic structure.
\end{abstract}

\maketitle

\section{Introduction}

Raman spectroscopy is a widely used non-destructive experimental technique to study the vibrational dynamics of molecules and condensed-phase systems \cite{DAS_2011_raman, Loudon_1964, Cardona_I_1983}.
Typically, first-principles calculations of the Raman spectrum are carried out via a first-order expansion of the dielectric susceptibility in terms of harmonic phonons \cite{Porezag_Pederson_1996, Skelton_Burton_Jackson_Oba_Parker_Walsh_2017, Bagheri_Komsa_2023}.
In order for the mode to have a nonzero (first-order) contribution to the Raman spectrum, it is necessary for the dielectric susceptibility to have a non-zero first-order derivative with respect to displacements along the phonon mode.
Which and how many phonon modes that have a non-zero contribution can be deduced from the symmetry of the crystal \cite{Cardona_II_1982}.

If symmetry forbids first-order scattering, such as in cubic crystals, inclusion of higher-order terms in the susceptibility expansion becomes a necessity.
Contributions from second-order terms were explored computationally early using a shell polarizability model \cite{Cowley_1964} and more recently through a direct expansion of the dielectric susceptibility fitted to \gls{dft} data \cite{Benshalom_Reuveni_Korobko_Yaffe_Hellman_2022}.

Alternatively, the \gls{md} simulation technique can be used to obtain the Raman spectrum \cite{McQuarrie_1976, Thomas_Brehm_Fligg_Vöhringer_Kirchner_2013}.
The time correlation function of the dielectric susceptibility of the system is then evaluated and the Raman spectrum is obtained by a Fourier transform
\cite{Putrino_Parrinello_2002, 
Medders_Paesani_2015, Sommers_Andrade_Zhang_Wang_Car_2020, Berger_Lv_Komsa_2023, Xu_Rosander_2023}.
The benefit of this approach is that it captures all anharmonicity as well as higher-order scattering of the Raman spectrum.
However, for the technique to be accurate and numerically efficient, both the force evaluation in the \gls{md} simulation and the computation of the dielectric susceptibility tensor must be done efficiently and accurately \cite{Berger_Komsa_2024, Xu_Rosander_2023}.

Perovskite oxides (\ce{ABO3}) constitute a broad and important class of multifunctional materials, known for their wide variety of chemical compositions and different structural distortions. 
Raman scattering has been extensively used to study the structural dynamics in these materials; see, e.g., Refs.~\cite{
Grzechnik_Wolf_McMillan_1997, Guennou_Bouvier_Kreisel_Machon_2010, Bartasyte_Margueron_Santiso_Hlinka_Simon_Gregora_Chaix-Pluchery_Kreisel_Jimenez_Weiss_2011, Hayashi_Nakamura_Ebina_2013, Toulouse_Amoroso_Xin_Veber_Hatnean_Balakrishnan_Maglione_Ghosez_Kreisel_Guennou_2019}.
At ambient pressure and high temperatures, many perovskite oxides are cubic, but upon cooling they undergo one or several phase transitions to structures with lower symmetries.
A rare exception is barium zirconate \ce{BaZrO3} (\gls{bzo}), which is claimed to remain cubic down to \qty{0}{\kelvin} \cite{Akbarzadeh_Kornev_Malibert_Bellaiche_Kiat_2005, Toulouse_Amoroso_Xin_Veber_Hatnean_Balakrishnan_Maglione_Ghosez_Kreisel_Guennou_2019, Perrichon_Jedvik_Granhed_Romanelli_Piovano_Lindman_Hyldgaard_Wahnström_Karlsson_2020, Rosander_Fransson_Milesi-Brault_Toulouse_Bourdarot_Piovano_Bossak_Guennou_Wahnström_2023}. 
However, its Raman spectrum is intense and shows sharp, well-defined features reminiscent of first-order scattering.
\cite{Chemarin_Rosman_Pagnier_Lucazeau_2000, Colomban_Slodczyk_2009, Karlsson_Matic_Knee_Ahmed_Eriksson_Börjesson_2008, Giannici_Shirpour_Longo_Martorana_Merkle_Maier_2011, Charrier-Cougoulic_Pagnier_Lucazeau_1999, Helal_Mori_Kojima_2016, Toulouse_Amoroso_Xin_Veber_Hatnean_Balakrishnan_Maglione_Ghosez_Kreisel_Guennou_2019, Gim_Sur_Lee_Lee_Moon_Oh_Kim_2022, Toulouse_Amoroso_Oliva_Xin_Bouvier_Fertey_Veber_Maglione_Ghosez_Kreisel_2022}.
This is unexpected, since first-order Raman peaks are not allowed by symmetry in a cubic system. 

It has been suggested that the rich Raman spectrum of \gls{bzo} is due to nanodomains
\cite{Chemarin_Rosman_Pagnier_Lucazeau_2000, Colomban_Slodczyk_2009}
or locally distorted regions \cite{Karlsson_Matic_Knee_Ahmed_Eriksson_Börjesson_2008, Giannici_Shirpour_Longo_Martorana_Merkle_Maier_2011}.
When increasing the pressure at room temperature, Chemarin \textit{et al.} \cite{Chemarin_Rosman_Pagnier_Lucazeau_2000} found that the intense Raman spectrum decreased in amplitude and tended to disappear when approaching \qty{9}{\giga\pascal}.
They argued that this was because the nanodomains were being forced to interact more strongly with increasing pressure, which eventually led to a continuous structure with long-range order.
Above \qty{9}{\giga\pascal}, a clear spectral change was observed, which they associated with a phase transition.

Another reason for the presence of sharp and well-defined features in the Raman spectrum at ambient pressure could be second-order Raman processes \cite{Charrier-Cougoulic_Pagnier_Lucazeau_1999}.
These processes are allowed by symmetry, but they are generally weaker in intensity compared with first-order scattering.
Two recent Raman studies on \gls{bzo} single crystals \cite{Toulouse_Amoroso_Xin_Veber_Hatnean_Balakrishnan_Maglione_Ghosez_Kreisel_Guennou_2019, Gim_Sur_Lee_Lee_Moon_Oh_Kim_2022} concluded that the Raman spectrum of \gls{bzo} at room temperature could be explained by second-order scattering.
Furthermore, it was also stated that local disorder could still play a role in the general intensity of the Raman spectrum \cite{Toulouse_Amoroso_Xin_Veber_Hatnean_Balakrishnan_Maglione_Ghosez_Kreisel_Guennou_2019}.

The pressure-dependent Raman scattering at room temperature was studied by Chemarin {\it et al.} \cite{Chemarin_Rosman_Pagnier_Lucazeau_2000}
using a polycrystalline sample.
They found two structural phase transitions, one at \qty{9}{\giga\pascal} and one at \qty{23}{\giga\pascal}.
Recently, Gim \textit{et al.} \cite{Gim_Sur_Lee_Lee_Moon_Oh_Kim_2022} and Toulouse \textit{et al.} \cite{Toulouse_Amoroso_Oliva_Xin_Bouvier_Fertey_Veber_Maglione_Ghosez_Kreisel_2022} also investigated the pressure dependence, now using single crystal samples.
Gim \textit{et al.} \cite{Gim_Sur_Lee_Lee_Moon_Oh_Kim_2022} found two transitions, from cubic to a rhombohedral phase at \qty{8.4}{\giga\pascal} and then from the rhombohedral to a tetragonal phase at \qty{19.2}{\giga\pascal}.
On the other hand, Toulouse \textit{et al.} \cite{Toulouse_Amoroso_Oliva_Xin_Bouvier_Fertey_Veber_Maglione_Ghosez_Kreisel_2022} found a single phase transition, from the cubic to a tetragonal phase at \qty{10}{\giga\pascal}.
They also discussed why a second phase transition is not expected for this system, at least for pressures below \qty{45}{\giga\pascal}.

In a previous paper \cite{FraRosErhWah2023} we developed a \gls{nep} model for the potential energy surface of \gls{bzo}. It was used to study \gls{bzo} at ambient conditions, but also its pressure dependence at room temperature, including the phase transition from the cubic to the tetragonal phase.

Here, we compute the Raman spectrum for \gls{bzo} via \gls{md} simulations. We use the same model for the potential energy surface as in Ref.~\cite{FraRosErhWah2023} and for the dielectric susceptibility we employ a \gls{dft} based model, recently developed using the \gls{tnep} framework \cite{Xu_Rosander_2023}. 

We consider the cubic phase of \gls{bzo} at ambient conditions, as well as under pressure including the transition to a tetragonal phase.
After correcting for the classical statistics in our \gls{md} approach we obtain excellent agreement with recent experimental Raman studies of single crystals of \gls{bzo} at ambient conditions.
The results are compared with the corresponding results from an expansion of the dielectric susceptibility to second order in terms of harmonic phonons.
This allows us to assign features and peaks in the Raman spectra to specific $\boldsymbol{q}$-points in reciprocal space.
Above the phase transition pressure, in the tetragonal phase, all first order Raman peaks are identified and compared with available experimental data.
Finally, the phase transition is scrutinized, including a study of the ``central peak'' in the Raman response.

\section{Theory}

\subsection{Raman spectrum}\label{sec:Raman}

We consider off-resonance Raman spectroscopy \cite{Born_Huang_1954, Cowley_1964}.
The frequency of the incoming $\omega_\mathrm{in}$ (and outgoing $\omega_\mathrm{out}$) light is assumed to be much larger compared to the phonon frequencies of the crystal and it is also assumed to be smaller than any electronic excitations in the material. 
The dielectric susceptibility tensor $\chi_{\alpha\beta}$ can then be determined in the so-called static ion-clamped limit \cite{Baroni_Resta_1986, Gonze_Lee_1997}.
Under these conditions the measured Raman intensity $I(\omega)$ is proportional to
\begin{equation}\label{eq:differential-cross-section}
    I(\omega) \propto \sum_{\alpha\beta\gamma\delta}\hat{n}^\mathrm{out}_\alpha\hat{n}^\mathrm{out}_\beta L_{\alpha\gamma\beta\delta}(\omega)\hat{n}^\mathrm{in}_\gamma\hat{n}^\mathrm{in}_\delta.
\end{equation}
Here, $\omega\equiv\omega_\mathrm{in}-\omega_\mathrm{out}$ is the Raman shift, and $\hat{\vec{n}}^\mathrm{in}$ and $\hat{\vec{n}}^{\mathrm{out}}$ are the polarization of the incoming and outgoing light, respectively, where $\alpha$, $\beta$, $\gamma$, as well as $\delta$ are Cartesian indices.
Furthermore, $\vec{L}(\omega)$ is the Raman lineshape, given by
\begin{equation}\label{eq:lineshape}
    L_{\alpha\gamma\beta\delta}(\omega) = 
    \frac{1}{2\pi} \int_{-\infty}^{\infty} \mathrm{d}t 
    \braket{ \chi_{\alpha\gamma}(t) \chi_{\beta\delta}(0) } 
    e^{-\mathrm{i}\omega t} ,   
\end{equation}
where the time-dependence of the dielectric susceptibility originates from the motion of the atoms in the crystal.
We note that the incoming and outgoing polarization of the light picks out the elements of the dielectric susceptibility, as indicated in \autoref{eq:differential-cross-section}.

The Raman scattering is usually discussed in terms of the order of the scattering.
The contribution of different orders can be analyzed by Taylor expanding the dielectric susceptibility in terms of the displacements of the atoms from their equilibrium positions,
\begin{equation}\label{eq:taylor}
    \begin{split}
        \chi_{\alpha\gamma} = \left(\chi_0\right)_{\alpha\gamma} + 
        &\sum_{i\epsilon}
        \left(R_{i}^{\epsilon}\right)_{\alpha\gamma}
        u_{i}^{\epsilon} + \\
        &\frac{1}{2}\sum_{ij\epsilon\eta} \left(R_{ij}^{\epsilon\eta}\right)_{\alpha\gamma}
        u_{i}^{\epsilon}u_{j}^{\eta} + \ldots\ ,
    \end{split}
\end{equation}
where $\left(R_{i}^{\epsilon}\right)_{\alpha\gamma}$, $\left(R_{ij}^{\epsilon\eta}\right)_{\alpha\gamma}$ and so on denote derivatives of the dielectric susceptibility with respect to atomic displacements $u$.
The indices $i$ and $j$ enumerate the atoms, while $\epsilon$ and $\eta$ denote Cartesian directions.
The order is then defined by how many atomic displacements that are involved in the correlation function for the Raman lineshape in \autoref{eq:lineshape}.
The first-order scattering contribution is obtained from the second term in the expansion of the dielectric susceptibility in \autoref{eq:taylor}, the second-order contribution from the third term and so on (see \autoref{sec:mode-expansion} for more details).
We note that in a cubic system the first order derivative is always zero due to the symmetry of the crystal, and there is thus no first-order scattering.

Most of our results for the Raman spectrum will be based on a direct evaluation of the time-correlation function in \autoref{eq:lineshape}, based on machine learning-accelerated \gls{md} simulations using the \gls{nep} \cite{FraRosErhWah2023} and \gls{tnep} models \cite{Xu_Rosander_2023}. 
These results are denoted as \textbf{MD}.
We will also show some results based on the expansion of \autoref{eq:taylor} to second order.
We use the same \gls{tnep} model to evaluate the derivatives, and the corresponding results will be denoted as \textbf{DSE}, the dielectric susceptibility expansion.

\section{Results}

\subsection{Room temperature spectrum} 

\begin{figure}[ht]
    \centering
    \includegraphics{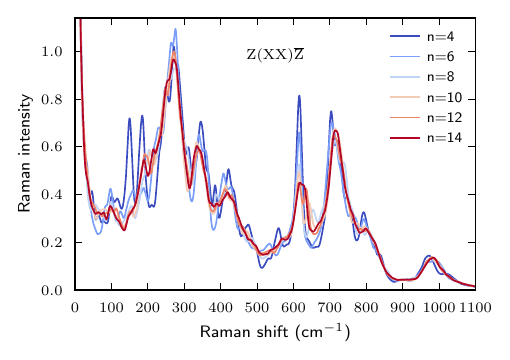}
    \caption{
        Molecular dynamics (MD) simulation of the Raman spectrum at \qty{300}{\kelvin} and \qty{0}{\giga\pascal}.
        Test of convergence with respect to the size of the supercell,
        which is given by (n$\times$n$\times$n) cubic primitive cells. The total number of atoms is 5n$^3$.
    }
    \label{fig:raman_spectrum_convergence}
\end{figure}

Consider first the system at room temperature (\qty{300}{\kelvin}) and ambient pressure (\qty{0}{\giga\pascal}).
The system is then cubic \cite{FraRosErhWah2023}.
We assume that the incoming and outgoing light are polarized along the same axis and that a crystal axis is aligned with the polarization of the light.
In Porto notation \cite{PortoNotation}, this is denoted as Z(XX)$\overbar{\text{Z}}$.
In \autoref{fig:raman_spectrum_convergence} we show the resulting Raman spectrum and its convergence as function of supercell size. 
It is clear that a size of at least \numproduct{10x10x10} cubic primitive cells is required to obtain well converged numerical results.

\begin{figure}[ht]
    \centering
    \includegraphics{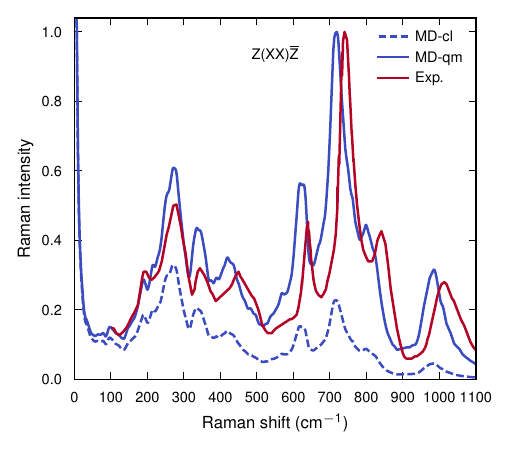}
    \caption{
        Molecular dynamics (MD) simulation of the Raman spectrum at \qty{300}{\kelvin} and \qty{0}{\giga\pascal} (MD-cl), with quantum corrections due to combination modes (MD-qm), compared with experiments at \qty{300}{\kelvin} (Exp.) from Ref.~\citenum{Toulouse_Amoroso_Xin_Veber_Hatnean_Balakrishnan_Maglione_Ghosez_Kreisel_Guennou_2019}. The experimental data are scaled such that the height of the peak around \qty{700}{\per\centi\meter} coincides between theory and experiments.
    }
    \label{fig:raman_simulated_experiment}
\end{figure}

Next, we compare with the experimental data from Toulouse {\it et al.} \cite{Toulouse_Amoroso_Xin_Veber_Hatnean_Balakrishnan_Maglione_Ghosez_Kreisel_Guennou_2019}.
Those data were obtained for a single crystal and the geometric setup was Z(XX)$\overbar{\text{Z}}$.
In \autoref{fig:raman_simulated_experiment} we show our \gls{md} results at \qty{300}{\kelvin} for the same geometric setup  as a dashed line (denoted MD-cl).
It is known that classical \gls{md} simulations underestimate the intensity of the true spectrum due to quantum effects present in the real system.
Those effects can be estimated by taking quantum statistics into account 
\cite{Cardona_II_1982}.
(For further details, see \autoref{ssec:qunatum-correction}).
In the present case, first-order contributions to the intensity are absent due to the cubic symmetry of the system. 
Second-order contributions can be either due to overtones or to the combination of two modes.
It has been stated that in solids, combination modes always dominate over overtones \cite{Cardona_II_1982}.
Therefore, we restrict ourselves to combination modes and in addition we assume a combination mode that consists of the sum of two different modes but with the same frequency (i.e., $\omega_1=\omega_2$ with $\omega=\omega_1+\omega_2$).
The correction for the classical treatment can then be written as
\begin{equation}\label{eq:rescaling}
    I_{\text{qm}}(\omega) = \left( \frac{\beta\hbar\omega/2}{1 - \exp{(-\beta\hbar\omega/2)}} \right)^2\ I_{\text{cl}}(\omega)\ .
\end{equation}
The result for $I_{\text{qm}}(\omega)$ is shown as a solid line (denoted MD-qm) in \autoref{fig:raman_simulated_experiment}. 

In the same figure we show the results from Toulouse \textit{et al.}    \cite{Toulouse_Amoroso_Xin_Veber_Hatnean_Balakrishnan_Maglione_Ghosez_Kreisel_Guennou_2019}, also at \qty{300}{\kelvin}.
The absolute intensity is unknown, and we have therefore scaled the experimental data such that the height of the peak around \qty{700}{\per\centi\meter} coincides between theory and experiments.
Our simulation agree very well with the experiments.
Taking the quantum statistics into account through the rescaling factor in \autoref{eq:rescaling} is important to obtain a qualitative agreement with experiments.
A slight red shift is present in our data compared to experiments, which we attribute to the underlying exchange-correlation functional (the van-der-Waals density functional with consistent exchange \cite{Dion_Rydberg_Schröder_Langreth_Lundqvist_2004,Berland_Hyldgaard_2014}), which is known to give slightly red-shifted vibrational frequencies for the present system (see Supp. Inf. in Ref.~\citenum{FraRosErhWah2023}).
For a similar comparison for the geometric setup 
Z(XY)$\overbar{\text{Z}}$, see \autoref{sfig:classic}b.

\subsection{Mode decomposition at room temperature}

The spectrum at \qty{300}{\kelvin} and \qty{0}{\giga\pascal} originates from higher-order Raman scattering processes and is likely dominated by second-order contributions. 
It is therefore tempting to try to identify the various peaks in the intensity with certain overtones and/or combination modes.
This can be done by making use of the \gls{dse} in \autoref{eq:taylor}, with the atomic displacements transformed to normal mode coordinates, see App.~\ref{sec:mode-expansion}.

To reduce the computational effort, we restrict the computations to a \numproduct{4x4x4} supercell (\num{320} atoms).
The result in \autoref{fig:raman_spectrum_convergence} shows that the intensity of this smaller supercell is not fully converged.
Nevertheless, the result from the smaller supercell contains sharp well-defined peaks, roughly with the correct positions and intensities.
Therefore, we conclude that the smaller supercell is sufficient for identifying possible overtones and/or combination modes in the spectrum.

\begin{figure}[ht]
    \centering
    \includegraphics{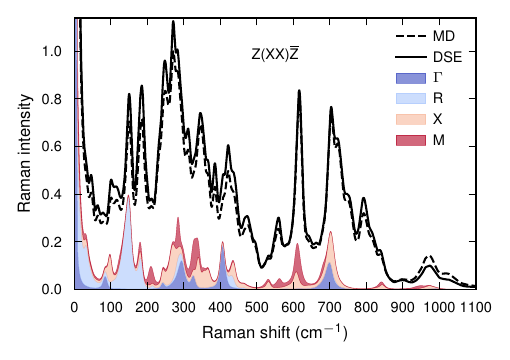}
    \caption{
        Raman spectrum based on the dielectric susceptibility expansion (DSE) to second order, compared with MD; both using a \numproduct{4x4x4} supercell.
        Contributions from different high symmetry $\boldsymbol{q}$-points to the total Raman spectrum are also shown.
        These contributions are shown cumulatively, \textit{i.e.} they are added up.
        The degeneracy of the high symmetry $\boldsymbol{q}$-points is included in plotted intensity.
    }
    \label{fig:raman_decomposition}
\end{figure}

In \autoref{fig:raman_decomposition} we show the result for the Raman spectrum using the \gls{dse} to second order.
The only term in the mode expansion that then contributes to the intensity in a cubic system is the one denoted by $L_{\alpha\gamma\beta\delta}^{\mathrm{II}}(\omega)$ in App.~\ref{sec:mode-expansion}.
In \autoref{fig:raman_decomposition} we also show the result using the \gls{md} method for the small \numproduct{4x4x4} supercell.
The result is nearly identical to the result using the expansion of the dielectric susceptibility.
The small difference may be due to the neglect of higher-order terms in the expansion or simply due to numerical/statistical noise.
In any case, this demonstrates that the Raman spectrum is dominated by second-order scattering.

The \gls{dse} result can then be decomposed into the contribution from different $\boldsymbol{q}$-points in the Brillouin zone, in total 64 points.
We set $\mathbf{k}=\mathbf{q}$ in \autoref{eq:Greens-second-second} and show the results for the intensities from the four high-symmetry points $\Gamma$, X, M, and R in \autoref{fig:raman_decomposition}.
These different contributions are shown cumulatively, \textit{i.e.} they are added up.
It is clear that the high symmetry points do not capture the full intensity of the Raman spectrum. 
In fact, most of the intensity comes from other parts of the Brillouin zone.
An integration over the whole Brillouin zone becomes necessary to obtain accurate features of the Raman spectrum.
A similar conclusion was made in Ref.~\citenum{Benshalom_Reuveni_Korobko_Yaffe_Hellman_2022} for \ce{NaCl}.

Nevertheless, there are significant contributions to specific peaks stemming from the high symmetry points, yet not significant enough to warrant an assignment.
It is clear that the R-point gives a substantial contribution to the spectrum below \qty{200}{\per\centi\meter}.
The peak at \qty{620}{\per\centi\meter} gets a contribution from both the M and X points, while the $\Gamma$ and X points contribute to the peak at \qty{700}{\per\centi\meter}. 
For the frequencies between \qty{200}{\per\centi\meter} and \qty{500}{\per\centi\meter} the individual contribution from the high symmetry points is less clear.

\subsection{Pressure dependence at room temperature}

\begin{figure}[ht]
    \centering
    \includegraphics{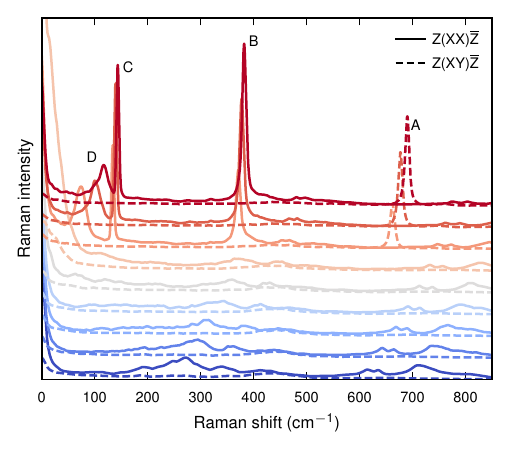}
    \caption{MD simulation of the Raman spectrum at
    \qty{300}{\kelvin} as function of pressure, from
    \qty{0}{\giga\pascal} to \qty{24}{\giga\pascal}
    with \qty{3}{\giga\pascal} increments. 
    At \qty{15}{\giga\pascal} a pronounced central peak is visible. The phase transition to the tetragonal phase occurs at about \qty{16.2}{\giga\pascal}.}
    \label{fig:raman_pressure}
\end{figure}

Next, we consider the pressure dependence of the Raman spectrum at room temperature (\qty{300}{\kelvin}).
Our system exhibits a phase transition from the cubic phase to the tetragonal phase at about \qty{16.2}{\giga\pascal} \cite{FraRosErhWah2023}.
In our setup, the elongated axis for the tetragonal system is oriented in the z-direction.

In \autoref{fig:raman_pressure} we present our result for the pressure dependence from \qtyrange{0}{24}{\giga\pascal}.
In the figure, we show the result for the setup  Z(XX)$\overbar{\text{Z}}$ and Z(XY)$\overbar{\text{Z}}$.
As we increase the pressure from \qty{0}{\giga\pascal}, the intensities of all peaks in the Raman spectrum are decreasing prior to the phase transition.
Hence, our results are fully in line with the experimentally measured spectra by Chemarin \textit{et al.}~\cite{Chemarin_Rosman_Pagnier_Lucazeau_2000} and by Toulouse \textit{et al.}~ \cite{Toulouse_Amoroso_Oliva_Xin_Bouvier_Fertey_Veber_Maglione_Ghosez_Kreisel_2022}.

As we further increase the pressure beyond the tetragonal phase transition, three distinct first order peaks appear in the spectrum.
We denote these by A, B, and C (see \autoref{fig:raman_pressure} and \autoref{tab:tetragonal_frequency}). 
These modes have previously been identified experimentally by Toulouse \textit{et al.}~ \cite{Toulouse_Amoroso_Oliva_Xin_Bouvier_Fertey_Veber_Maglione_Ghosez_Kreisel_2022}.

Close to, but below, the phase transition the quasi-elastic line broadens and increases substantially in intensity to a broad ``central Raman peak'', which is clearly seen in \autoref{fig:raman_pressure} at \qty{15}{\giga\pascal}
(see also \autoref{sfig:raman-pressure}a).
This behavior is reminiscent of the over-damped tilt-mode visible in the dynamical structure factor of \gls{bzo} close to the phase transition \cite{FraRosErhWah2023}.

\begin{table*}[ht]
    \centering
    \begin{tabular}{lccccccc}
    \toprule
                                         & A               & B1                      & B2                      & C1                      & C2                      & D                       & E                         \\
    \midrule
        Frequency $\omega_0$ (\unit{\per\centi\meter}) & 677             & 378                     & 369                     & 139                     & 136                     & 103                     & 64                        \\
        Damping $\Gamma$   (\unit{\per\centi\meter})   &  9.1            &  10.2                   &  12.0                   &   5.4                   &   5.8                   &  27.8                   & 20.5                      \\
        Slope     (\unit{\per\centi\meter}/GPa)        &  4.7            &   1.8                   &   0.8                   &   1.6                   &   1.2                   &   7.0                   & 2.8                       \\
        \multirow{3}*{Visibility}&Z(XY)$\overbar{\text{Z}}$&Z(XX)$\overbar{\text{Z}}$&Y(XZ)$\overbar{\text{Y}}$&Z(XX)$\overbar{\text{Z}}$&Y(XZ)$\overbar{\text{Y}}$&Z(XX)$\overbar{\text{Z}}$&Y(XZ)$\overbar{\text{Y}}$  \\
                                 &                         &                         &                         &                         &                         &Z(YY)$\overbar{\text{Z}}$&X(YZ)$\overbar{\text{X}}$  \\
                                 &                         &                         &                         &                         &                         &                         &Y(ZZ)$\overbar{\text{Y}}$  \\
        Raman lineshape                  &   xyxy          &    xxxx                 &  xzxz                   &  xxxx                   & xzxz                    &   xxxx,yyyy             &  xzxz,yzyz,zzzz           \\
        \bottomrule
    \end{tabular}
    \caption{
        Frequency, damping and slope for the first-order active Raman modes in the tetragonal phase at \qty{300}{\kelvin} and $\qty{21}{\giga\pascal}$.
        The frequency $\omega_0$ and damping $\Gamma$ are obtained by fitting to a damped harmonic oscillator model $I(\omega) \propto 2 \Gamma \omega_0^2/[(\omega^2 - \omega_0^2)^2 + (\Gamma\omega)^2]$ \cite{Fransson_Slabanja_Erhart_Wahnström_2021}.
        The pressure dependence of the peak position, i.e., the slope, is then determined by a finite difference approximation.
        For each mode the corresponding Porto notation is given as well as the indices $\alpha\gamma\beta\delta$ for the Raman lineshape.
    }
    \label{tab:tetragonal_frequency}
\end{table*}

\begin{figure}[ht]
    \centering
    \includegraphics{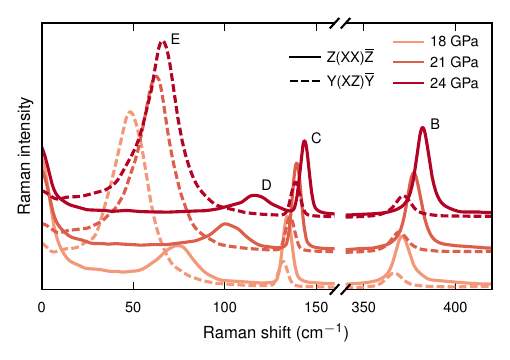}
    \caption{
        Results from MD simulations of the Raman spectra at \qty{300}{\kelvin} and at three different pressure in the tetragonal phase, showcasing under which experimental setups the first order active phonon modes are visible.
    }
    \label{fig:raman_pressure_wide_range}
\end{figure}

The high frequency A mode corresponds to Jahn-Teller-like distortions of oxygen octahedra \cite{Toulouse_Amoroso_Oliva_Xin_Bouvier_Fertey_Veber_Maglione_Ghosez_Kreisel_2022}.
Its pressure dependence is significant with a slope of about \qty{4.7}{\per\centi\meter\per\giga\pascal} and is only present in the polarization setup Z(XY)$\overbar{\text{Z}}$.
The intensity is initially quite small, compared with mode B and C, but its intensity increases more with increasing pressure compared with B and C.
This mode, A, is clearly visible in Chemarin \textit{et al.} \cite{Chemarin_Rosman_Pagnier_Lucazeau_2000} and also in Gim \textit{et al.} \cite{Gim_Sur_Lee_Lee_Moon_Oh_Kim_2022}, albeit less pronounced.
The pressure dependence of the peak position is similar in these two experimental studies and in line with what we find here.
In Ref.~\citenum{Toulouse_Amoroso_Oliva_Xin_Bouvier_Fertey_Veber_Maglione_Ghosez_Kreisel_2022} the mode is very faint and the authors state that this is most likely a consequence of the experimental setup and related to the orientation of the single crystal.

The other two peaks, B and C, both consist of two closely overlapping bands and stem from the lifting of degenerate modes in the cubic cell (see \autoref{fig:raman_pressure_wide_range}).
A large peak is visible in Z(XX)$\overbar{\text{Z}}$ and a smaller peak is visible in Y(XZ)$\overbar{\text{Y}}$, and where the smaller peak has a slightly lower frequency.
The B mode is due to oxygen octahedra shearing modes and the C mode to antiparallel barium motion \cite{Toulouse_Amoroso_Oliva_Xin_Bouvier_Fertey_Veber_Maglione_Ghosez_Kreisel_2022}.
The splitting of the B mode is visible in spectra of Chemarin \textit{et al.} \cite{Chemarin_Rosman_Pagnier_Lucazeau_2000} and Gim \textit{et al.} \cite{Gim_Sur_Lee_Lee_Moon_Oh_Kim_2022}, but only at considerably higher pressures than the phase transition pressure.
Therefore, they both concluded that a second phase transition occurs.
Here we show that the splitting is present already at the phase transition to the tetragonal phase and no further phase transition occurs.
Toulouse \textit{et al.} \cite{Toulouse_Amoroso_Oliva_Xin_Bouvier_Fertey_Veber_Maglione_Ghosez_Kreisel_2022} could not resolve any splitting for the B mode but noted that the C mode has an asymmetric profile at higher pressure, which may indicate the presence of two closely overlapping bands.

The Raman spectrum also contains two soft modes, here denoted D and E, and visible in the setup Z(XX)$\overbar{\text{Z}}$ and Y(XZ)$\overbar{\text{Y}}$, respectively.
They are due to the tilt mode of the octahedra, and, in particular, the position of the D mode shows a strong dependence on the pressure.
Both Gim \textit{et al.} \cite{Gim_Sur_Lee_Lee_Moon_Oh_Kim_2022} and Toulouse \textit{et al.} \cite{Toulouse_Amoroso_Oliva_Xin_Bouvier_Fertey_Veber_Maglione_Ghosez_Kreisel_2022} detected these two modes.
They also found that the mode with higher frequency (the D-mode) shows a stronger pressure dependence.
In the study by Toulouse \textit{et al.} \cite{Toulouse_Amoroso_Oliva_Xin_Bouvier_Fertey_Veber_Maglione_Ghosez_Kreisel_2022} the soft mode with lower frequency (the E mode) is lost in the quasi-elastic line at low pressures.

\subsection{Phase transition}

The phase transition from the cubic to the tetragonal phase is driven by tilting of the \ce{ZrO6} octahedra \cite{FraRosErhWah2023, Samara1975, Megaw1971, Angel2005, Tohei2005}.
The corresponding phonon mode is located at the R-point in the phonon dispersion relation for the cubic structure.
It is therefore instructive to consider the dynamical behavior of the phonon mode coordinate $Q(t)$ for the R-tilt mode.
The latter can be obtained from \gls{md} simulations by projecting the atomic displacements $\mathbf{u}(t)$ on the supercell eigenvector, as done in Ref.~\cite{FraRosErhWah2023},
\begin{align}
  Q_\lambda(t) = \braket{\ \bold{e}_\lambda \mid \bold{u}(t)\ }
\end{align}
where $\mathbf{e}_\lambda$ is the supercell eigenvector for a given mode $\lambda$.
In the cubic phase, the mode exhibits threefold degeneracy along the Cartesian directions, while in the tetragonal phase, this degeneracy is broken, making one direction symmetrically distinct from the other two.
We denote these directions as z and xy, respectively.

\begin{figure}[ht]
    \centering
    \includegraphics{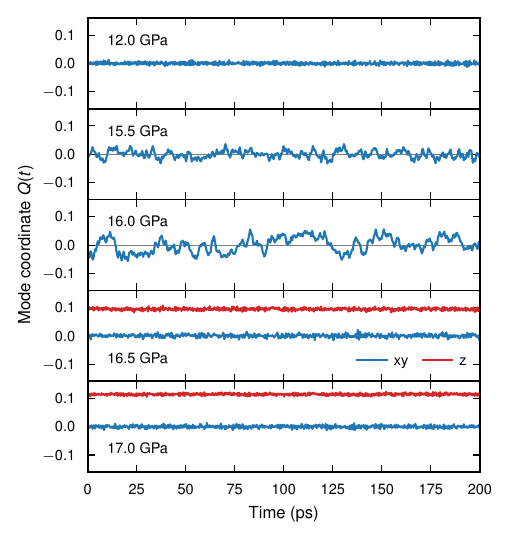}
    \caption{
        Mode coordinate in time, $Q(t)$, of the R-mode at \qty{300}{\kelvin} and at five different pressures; \qtylist[list-units=single,range-phrase=, ]{12.0;15.5;16.0;16.5;17.0}{\giga\pascal}.
    }
    \label{fig:modes_in_time}
\end{figure}

In \autoref{fig:modes_in_time} we show the time evolution of the phonon mode coordinates $Q(t)$ at five different pressures, three below the phase transition and two above.
(For the time evolution at some other pressures, see \autoref{sfig:modes_in_time}.)
At \qty{12.0}{\giga\pascal} the mode coordinate shows quite small and regular oscillations. 
When the pressure is increased and is approaching the phase transition, the oscillations become larger, much more irregular, and the time-scale of the motion is slowing down.
Above the phase transition the oscillations again become smaller, faster and more regular and the degeneracy is broken.

\begin{figure}[ht]
   \centering
   \includegraphics{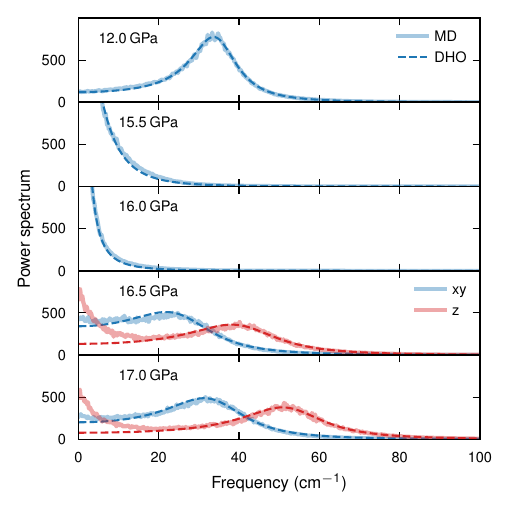}
    \caption{
        Power spectrum of the mode coordinate of the R-mode at \qty{300}{\kelvin} and at five different pressures; \qtylist[list-units=single,range-phrase=, ]{12.0;15.5;16.0;16.5;17.0}{\giga\pascal}. The results from the MD simulation are compared with the corresponding results using the damped harmonic oscillator (DHO) model.
    }
    \label{fig:Power_spectrum}
\end{figure}

Consider next the spectral properties of $Q_\lambda$.
Its power spectrum can be obtained from the Fourier transform of the auto-correlation function of $Q_\lambda(t)$, according to
\begin{equation}
    P_\lambda(\omega) = 
    \int \textrm{d}t\ e^{-i\omega t}\left < Q_\lambda(t+t') Q_\lambda(t') \right >
\end{equation}
This is shown in \autoref{fig:Power_spectrum} and for some further pressures in \autoref{sfig:mode_powerspectrum}.
At \qty{12.0}{\giga\pascal} the spectrum shows a broad peak located around \qty{35}{\per\centi\meter}. 
When increasing the pressure the frequency softens, the motion becomes overdamped and the spectrum instead develops a central peak, which increases in height when approaching the phase transition. 
Above the phase transition, where the system is tetragonal, the spectra comprise two broad peaks.
These correspond to the modes here denoted D and E in the Raman spectra (\autoref{fig:raman_pressure_wide_range}).
In addition, an intensity at low frequencies for the z-component is present, decreasing slightly when increasing the pressure. The corresponding auto-correlation function in time $C(t)= \braket{Q_\lambda(t+t')Q_\lambda(t')}$ is shown in \autoref{sfig:mode_acfs}. When approaching the phase transition from below, $C(t)$ decays exponentially, with a decay time that approaches infinity.

\begin{figure}[ht]
   \centering
    \includegraphics{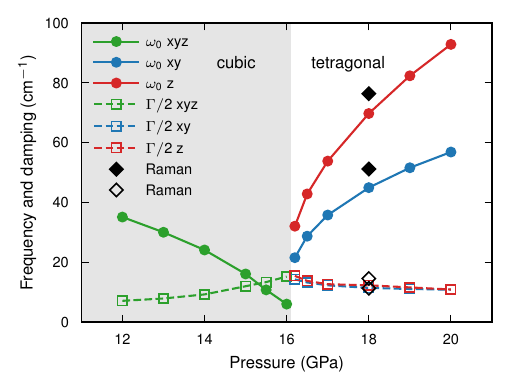}
    \caption{
        Frequency, $\omega_0$, and damping, $\Gamma/2$, for the R-mode, obtained by fitting to the damped harmonic oscillator (DHO) model, as a function of pressure at \qty{300}{\kelvin}.
        For comparison, the results from the Raman simulation are shown as black symbols.
    }
    \label{fig:frequencyVpressure}
\end{figure}

It is instructive to fit our results for the power spectrum to a \gls{dho} model (see \autoref{ssec:langevin} and 
Refs. \cite{Fransson_Rosander_Eriksson_Rahm_Tadano_Erhart_2023} and \cite{Fransson_Slabanja_Erhart_Wahnström_2021}).
That model is defined by two parameters, the natural frequency 
$\omega_0$ and the damping $\Gamma$.
In \autoref{fig:Power_spectrum} we show the fitted result as dashed lines. 
The obtained frequencies, $\omega_0$, and damping $\Gamma / 2$, are shown in \autoref{fig:frequencyVpressure}.
The \gls{dho} 
model describes the spectra in \autoref{fig:Power_spectrum} very well, except for the central peak observed in the tetragonal phase, mainly in the z-direction.
The obtained pressure dependence on the frequency $\omega_0$ in \autoref{fig:frequencyVpressure} indicates a continuous phase transition.
However, we note that close to a continuous phase transition, both the length and time-scale of the tilt mode diverges, rendering it difficult to converge the power spectra and thus leading to larger uncertainties close to the phase transition \cite{Fransson_Rosander_Eriksson_Rahm_Tadano_Erhart_2023}.

In the tetragonal phase, the power spectrum in the z-direction shows
both a broad peak at a finite frequency, an oscillatory peak, and an increased intensity at low frequencies, a central peak.
A central peak appears in the spectrum when the damping is large, 
$\Gamma/2 > \omega_0$, while an oscillatory peak appears when the damping is small, $\Gamma/2 < \omega_0$.
From symmetry, we expect the potential energy function for the phonon mode coordinate in the z-direction to be asymmetric.
The damping could also vary as function of distance.
In \autoref{ssec:langevin} we show that a harmonic well with distance dependent damping can show a power spectrum with both an oscillatory peak and a central peak. The same qualitative behavior can also be obtained using an asymmetric Morse potential together a constant, distance-independent damping. For further details, see \autoref{ssec:langevin}.

Let us now consider the Raman spectra in \autoref{fig:raman_pressure}.
As already stated, close to, but below the phase transition the quasi-elastic line broadens and increases substantially in intensity to a broad ``central Raman peak'' (see \autoref{fig:raman_pressure} and \autoref{sfig:raman-pressure}a).
At \qty{15}{\giga\pascal} the intensity increases rapidly for frequencies below $\sim\,\qty{80}{\per\centi\meter}$.
This is the most apparent signature of the onset of the phase transition.
The emergence of such a ``central Raman peak'' has been discussed for other perovskite materials 
\cite{Yaffe_Guo_Tan_Egger_Hull_Stoumpos_Zheng_Heinz_Kronik_Kanatzidis_2017, Seiler_Palato_Sonnichsen_Baker_Socie_Strandell_Kambhampati_2019, Gao_Yadgarov_Sharma_Korobko_McCall_Fabini_Stoumpos_Kanatzidis_Rappe_Yaffe_2021,  Hehlen_Bourges_Rufflé_Clément_Vialla_Ferreira_Ecolivet_Paofai_Cordier_Katan_2022, Berger_Komsa_2024, 
Lim_Righetto_Yan_Patel_Siday_Putland_McCall_Sirtl_Kominko_Peng_2024}. In the case of \gls{bzo}, we find here that the emergence of the ``central Raman peak'' is due to the overdamped behavior of the tilt mode close to, but below, the phase transition.

\begin{figure}[ht]
   \centering
   \includegraphics{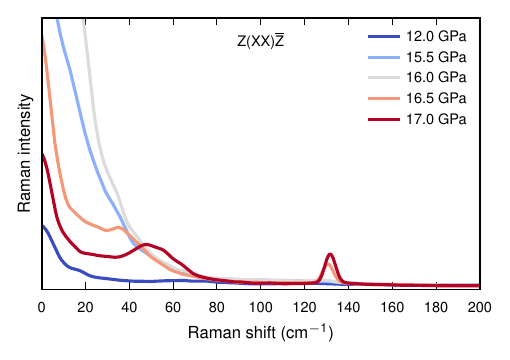}
    \caption{
        Raman spectra at \qty{300}{\kelvin} as a function of pressure, at \qtylist[list-units=single,range-phrase=, ]{12.0;15.5;16.0;16.5;17.0}{\giga\pascal}.
        The phase transition to the tetragonal phase occurs at about \SI{16.2}{GPa}
    }
    \label{fig:raman_pressure_range_zoom}
\end{figure}

In \autoref{fig:raman_pressure_range_zoom} we show the behavior of the Raman intensity around the phase transition temperature in more detail (see also \autoref{sfig:raman-pressure}b).
In the cubic phase the peak grows substantially close to the phase transition, similar to the overdamped power spectra in \autoref{fig:Power_spectrum},
despite that first-order scattering is forbidden in a cubic crystal.
However, second order scattering can be obtained with \autoref{eq:raman-2} together with Wick's theorem in \autoref{eq:wicks}, where the overtone of the tilt mode is given as a convolution of its own power spectral density.
Notably, there is an asymmetry in the phase transition, i.e., the pronounced central peak is only present in the cubic phase.
This is due to that the tilt mode quickly stiffens with pressure, and the mode becomes underdamped.

It is interesting to compare the low frequency dynamics of \gls{bzo}, an oxide perovskite, with the halide perovskites, which often exhibit a low-frequency Raman response, and thus a central Raman peak.
While the origin of this peak has been discussed extensively \cite{Yaffe_Guo_Tan_Egger_Hull_Stoumpos_Zheng_Heinz_Kronik_Kanatzidis_2017, Seiler_Palato_Sonnichsen_Baker_Socie_Strandell_Kambhampati_2019, Gao_Yadgarov_Sharma_Korobko_McCall_Fabini_Stoumpos_Kanatzidis_Rappe_Yaffe_2021, Hehlen_Bourges_Rufflé_Clément_Vialla_Ferreira_Ecolivet_Paofai_Cordier_Katan_2022, Berger_Komsa_2024,Lim_Righetto_Yan_Patel_Siday_Putland_McCall_Sirtl_Kominko_Peng_2024,Lanigan_2021,Weadock_2023,Songvilay_2019}, no consensus has yet been reached.
Also for the halide perovskites the central Raman peak can be understood from heavily overdamped tilt-modes that give rise to correlations on a very long time-scale and hence a narrow central Raman peak in the spectrum \cite{Fransson_Rosander_Eriksson_Rahm_Tadano_Erhart_2023}.
However, in the halide perovskites, both the R-point and M-point modes can become overdamped, not only the R-point mode, which indicates two-dimensional octahedral fluctuations \cite{Lanigan_2021,Fransson_Rosander_Eriksson_Rahm_Tadano_Erhart_2023}.
Furthermore, halide perovskites are often softer compared to the oxide perovskites and the effect is therefore more pronounced in the halide perovskites, but here we show that a central Raman peak can also appear in a ``hard'' oxide perovskite close to a phase transition.

\section{Conclusions}

We have computationally determined the Raman spectrum for \gls{bzo} by directly evaluating the dynamic auto-correlation function of the dielectric susceptibility tensor of the system using classical \gls{md} simulations.
To this end, we have used machine-learned models from the literature for the potential energy surface \cite{FraRosErhWah2023} as well as for the dielectric susceptibility tensor \cite{Xu_Rosander_2023}.

It has been established that at room temperature and ambient pressure \gls{bzo} is cubic, indicating that the Raman spectrum should lack sharp features.
Nevertheless, sharp features reminiscent of first-order scattering have been observed experimentally.
We also observe such features in our simulations and when correcting for the classical sampling in our approach, the simulated and experimental spectra are in excellent agreement.
The slight redshift of our spectrum can be attributed to the exchange-correlation functional used for the construction of the potential energy model, which is known to yield a slightly too soft response.
We can therefore conclude that the sharp features present in the experimental spectrum are due to higher-order scattering processes.

We have also determined the Raman spectrum by expanding the dielectric susceptibility tensor in terms of the phonon mode coordinates.
As we find that an expansion to second order gives almost indistinguishable results compared to the full model, we can conclude that the scattering is dominated by second-order effects.
The \gls{dse} then allows us to disentangle the contributions from different points in the Brillouin zone.
There are significant contributions to specific peaks stemming from the high symmetry points, but not significant enough to warrant any assignment.
Therefore, it becomes important to account for all points in the Brillouin zone to correctly capture the full Raman spectrum.

When the pressure is increased, the intensity of the peaks in the cubic structure is reduced, which is in line with experiments.
In light of this finding and based on our previous study \cite{FraRosErhWah2023}, we find no evidence of nano-domains as suggested by Chemarin \textit{et al.} \cite{Chemarin_Rosman_Pagnier_Lucazeau_2000}.

Additionally, slightly below the phase transition pressure, a broad central peak appears which is reminiscent of the behavior of the overdamped tilt-mode, which is a strong indication of the phase transition \cite{FraRosErhWah2023, Fransson_Rosander_Eriksson_Rahm_Tadano_Erhart_2023}.
Such a broad ``central Raman peak'' has been extensively discussed in connection to halide perovskites \cite{Lim_Righetto_Yan_Patel_Siday_Putland_McCall_Sirtl_Kominko_Peng_2024}.
Here, we show that it can also be present in a ``hard'' oxide perovskite and it originates from the highly overdamped R-tilt mode in the cubic structure.

Above the phase transition pressure, in the tetragonal phase, all first order Raman active peaks are identified.
Furthermore, we note that the splitting of the peaks at higher pressure seen experimentally
\cite{Chemarin_Rosman_Pagnier_Lucazeau_2000, Gim_Sur_Lee_Lee_Moon_Oh_Kim_2022, Toulouse_Amoroso_Oliva_Xin_Bouvier_Fertey_Veber_Maglione_Ghosez_Kreisel_2022}, is not due to a second phase transition.
Instead, this splitting is caused by the slightly different pressure dependence of the underlying phonon modes, and thus, the difference in frequency between the two modes increases as we increase the pressure.
Our simulation consolidates the analysis of Toulouse \textit{et al.} \cite{Toulouse_Amoroso_Oliva_Xin_Bouvier_Fertey_Veber_Maglione_Ghosez_Kreisel_2022} of the Raman spectrum in the tetragonal cell; there is only one phase transition for \gls{bzo}, at least up to \qty{45}{\giga\pascal}.

The present study shows that Raman spectra computed by \gls{md} simulations with machine-learned models for the potential energy surface and for the dielectric susceptibility can provide detailed and crucial information about the dynamics of the lattice vibrations and their impact on the Raman spectrum.

\section{Acknowledgments}
Funding from the Swedish Energy Agency (grant No. 45410-1), the Swedish Research Council (Nos. 2018-06482, 2020-04935 and 2021-05072) and the Excellence Initiative Nano at Chalmers is gratefully acknowledged. 
The computations were enabled by resources provided by the National Academic Infrastructure for Supercomputing in Sweden (NAISS) at C3SE, NSC, and PDC partially funded by the Swedish Research Council through grant agreements no. 2022-06725 and no. 2018-05973 as well as the Berzelius resource provided by the Knut and Alice Wallenberg Foundation at NSC. Computational resources provided by Chalmers e-commons are also acknowledged.

\appendix

\section{Mode decomposition} \label{sec:mode-expansion}

\subsection{Dielectric susceptibility expansion (DSE)}

Start by expressing the displacements $u_{i}^{\epsilon} \equiv u(i,\epsilon)$ in \autoref{eq:taylor} in terms of phonon operators $Q_{\mathbf{q}\nu}$ according to
\begin{equation}
    u(i, \epsilon) = \sum_{\mathbf{q}\nu}c_{\mathbf{q}\nu}(i, \epsilon)Q_{\mathbf{q}\nu}
\end{equation}
with,
\begin{equation}
    c_{\mathbf{q}\nu}(i, \epsilon) = \frac{\mathrm{e}^{\mathrm{i}\mathbf{q}\cdot\mathbf{r}(i)}}{\sqrt{Nm_i}}W_{\mathbf{q}\nu}(i, \epsilon),
\end{equation}
\begin{equation}
    Q_{\mathbf{q}\nu} = \sqrt{\frac{\hbar}{2\omega_{\mathbf{q}\nu}}}(a_{\mathbf{q}\nu} + a_{-\mathbf{q}\nu}^\dagger)
\end{equation}
where $m_i$ is the mass of atom $i$, $\mathbf{r}(i)$ its equilibrium position in the supercell and $N$ is the number of unit cells. Latin letters denote atoms in the supercell and Greek letters indicate a Cartesian direction.
Further, $\omega_{\mathbf{q}\nu}$ is the phonon frequency, $W_{\mathbf{q}\nu}$ the phonon eigenvector, and $a_{\mathbf{q}\nu}$ and $a^\dagger_{\mathbf-{q}\nu}$ are the creation and annihilation operators. $\mathbf{q}$ is used to denote a point in the first Brillouin zone and $\nu$ the corresponding branch index.

The reformulation of the atomic displacements in terms of phonon mode coordinates allows us to rewrite the Taylor expansion of the dielectric susceptibility in \autoref{eq:taylor} in terms of phonon mode coordinates \cite{Cardona_II_1982,Weber_Merlin_2000,Born_Huang_1954},
\begin{equation}\label{eq:mode-taylor}
    \begin{split}
        \boldsymbol{\chi} = \boldsymbol{\chi}_0 + 
        &\sum_{\mathbf{q}\nu}
        \tilde{\mathbf{R}}_{\mathbf{q}}^{\nu}
        Q_{\mathbf{q}\nu} \delta_{\mathbf{q},\boldsymbol{\Gamma}} \\
        &+ \frac{1}{2}\sum_{\substack{\mathbf{q}\nu\\\mathbf{k}\mu}} \tilde{\mathbf{R}}_{\mathbf{q};\mathbf{k}}^{\nu;\mu}
        Q_{\mathbf{q}\nu}Q_{\mathbf{k}\mu}\delta_{\mathbf{q},\mathbf{k}} + \ldots\ ,
    \end{split}
\end{equation}
where $\tilde{\mathbf{R}}$ is the Raman tensor for the respective order.
The restriction imposed by the two $\delta$-functions in \autoref{eq:mode-taylor} is a consequence of the invariance of the crystal against a rigid body translation.

The Raman tensor for the first and second order is then given by,
\begin{alignat*}{2}
        \left(\tilde{\mathrm{R}}_{\mathbf{q}}^{\nu}\right)_{\alpha\gamma} = \sum_{i,\epsilon}                                  &\frac{\partial\chi_{\alpha\gamma}}{\partial u(i, \epsilon)}c_{\mathbf{q}\nu}(i, \epsilon) W_{\mathbf{q}\nu}(i, \epsilon), \\
        \left(\mathrm{\tilde{R}}_{\mathbf{q};\mathbf{k}}^{\nu;\mu}\right)_{\alpha\gamma}=\sum_{\substack{i,\epsilon\\j,\zeta}} &\frac{\partial^2\chi_{\alpha\gamma}}{\partial u(i, \epsilon)\partial u(j,\zeta)}c_{\mathbf{q}\nu}(i, \epsilon)c_{\mathbf{k}\mu}(j,\zeta) \\
        &W_{\mathbf{q}\nu}(i, \epsilon)W_{\mathbf{k}\mu}(i, \zeta),
\end{alignat*}
As we note in \autoref{sec:Raman}, the first order derivative with respect to atomic displacements is zero for a cubic system, therefore, the first order Raman intensities expressed in phonon coordinates will consequentially also be zero.

Inserting the expansion of the dielectric susceptibility, \autoref{eq:mode-taylor}, into the quantum mechanical expression for the Raman lineshape (cf. \autoref{eq:lineshape}), 
\begin{equation*}
    L_{\alpha\gamma\beta\delta}(\omega) = 
    \frac{1}{2\pi} \int_{-\infty}^{\infty} \mathrm{d}t 
    \braket{ \chi_{\alpha\gamma}(t) \chi^\dagger_{\beta\delta}(0)} 
    e^{-\mathrm{i}\omega t},
\end{equation*}
leads to an expansion of the lineshape in terms of Fourier transformed phonon Green's functions and Raman intensities.
\begin{equation}
    \begin{split}
            L_{\alpha\gamma\beta\delta}^{\mathrm{I}}(\omega) = 
            \frac{1}{4\pi}\sum_{\nu\mu}
            &(\tilde{R}_{\boldsymbol{\Gamma}}^{\nu})_{\alpha\gamma}
            (\tilde{R}_{\boldsymbol{\Gamma}}^{\mu})_{\beta\delta}^\dagger
            \tilde{G}_{\boldsymbol{\Gamma}}^{\nu;\mu}(\omega),
    \end{split}
\end{equation}
\begin{equation}\label{eq:raman-2}
    \begin{split}
            L_{\alpha\gamma\beta\delta}^{\mathrm{II}}(\omega) = 
            \frac{1}{8\pi}\sum_{\substack{\mathbf{q}\mathbf{k}\\\nu\mu\nu'\mu'}}
            &(\tilde{R}_{\mathbf{q};-\mathbf{q}}^{\nu;\mu})_{\alpha\gamma}
            (\tilde{R}_{\mathbf{k};-\mathbf{k}}^{\nu';\mu'})_{\beta\delta}^\dagger
            \tilde{G}_{\mathbf{q};\mathbf{k}}^{\nu\mu;\nu'\mu'}(\omega),
    \end{split}
\end{equation}
\begin{equation}
    \begin{split}
            L_{\alpha\gamma\beta\delta}^{\mathrm{III}}(\omega) = 
            \frac{1}{4\pi}\sum_{\substack{\mathbf{k}\\\nu\mu\nu'\mu'}}
            &(\tilde{R}_{\boldsymbol{\Gamma}}^{\nu})_{\alpha\gamma}
            (\tilde{R}_{\mathbf{k};-\mathbf{k}}^{\nu';\mu'})_{\beta\delta}^\dagger
            \tilde{G}_{\boldsymbol{\Gamma};\mathbf{k}}^{\nu;\nu'\mu'}(\omega) + \\
            &(\tilde{R}_{\mathbf{k};-\mathbf{k}}^{\nu;\mu})_{\alpha\gamma}
            (\tilde{R}_{\boldsymbol{\Gamma}}^{\nu'})_{\beta\delta}^\dagger
            \tilde{G}_{\mathbf{k};\boldsymbol{\Gamma}}^{\nu\mu;\nu'}(\omega)
    \end{split}
\end{equation}
The phonon Green's functions in the time domain are defined as,
\begin{equation}\label{eq:Greens-first}
    G_{\boldsymbol{\Gamma}}^{\nu;\mu}(t) = \braket{Q_{\boldsymbol{\Gamma}\nu}(t)Q_{\boldsymbol{\Gamma}\mu}^{\dagger}(0)} \propto \braket{A_{\boldsymbol{\Gamma}\nu}(t)A_{\boldsymbol{\Gamma}\mu}^{\dagger}(0)}
\end{equation}
\begin{equation}\label{eq:Greens-second-second}
    G_{\mathbf{q};\mathbf{k}}^{\nu\mu;\nu'\mu'}(t) \propto \braket{A_{\mathbf{q}\nu}(t)A_{-\mathbf{q}\mu}(t)A_{\mathbf{k}\nu'}^{\dagger}(0)A_{-\mathbf{k}\mu'}^{\dagger}(0)}.
\end{equation}
\begin{equation}\label{eq:Greens-first-second}
    G_{\boldsymbol{\Gamma};\mathbf{q}}^{\nu;\nu'\mu'}(t) \propto \braket{A_{\boldsymbol{\Gamma}\nu}(t)A_{\mathbf{q}\nu'}^{\dagger}(0)A_{-\mathbf{q}\mu'}^{\dagger}(0)}
\end{equation}
\begin{equation}\label{eq:Greens-second-first}
    G_{\mathbf{q};\boldsymbol{\Gamma}}^{\nu\mu;\mu'}(t) \propto \braket{A_{\mathbf{q}\nu}(t)A_{-\mathbf{q}\mu}(t)A_{\boldsymbol{\Gamma}\mu'}^{\dagger}(0)}
\end{equation}
where $A_{\mathbf{q}\nu}(t) = \text{exp}(t H /\hbar)A_{\mathbf{q}\nu}\text{exp}(-t H /\hbar)$ and $A_{\mathbf{q}\nu} = a_{\mathbf{q}\nu} + a_{\mathbf{-q}\nu}^{\dagger}$, i.e., this is the phonon displacement operator in the Heisenberg picture.

For harmonic systems, the modes don't mix, this means that only $G_{\boldsymbol{\Gamma}}^{\nu;\mu}(t)$ with $\nu=\mu$ and $G_{\mathbf{q};\mathbf{k}}^{\nu\nu';\mu\mu'}(t)$ with $\mathbf{q} = \mathbf{k}$, $\nu=\nu'$ and $\mu=\mu'$ is non-zero.
The first term would then correspond to first order Raman scattering and the second term would be second order Raman scattering.
For the second order, $\nu=\mu$ is referred to as overtones whereas, $\nu\neq\mu$ is referred to as combination modes.
For anharmonic systems, the other terms do not necessarily vanish, instead they will contribute to the one-phonon peaks but they decay rapidly away from these peaks \cite{Benshalom_Reuveni_Korobko_Yaffe_Hellman_2022}.

We make a {\it classical} approximation and obtain these Greens functions by projecting the atomic displacements on the mode coordinates during the \gls{md} simulation, see e.g. \cite{SunSheAll2010,CarTogTan2017,RohLiLuoHen2022,Berger_Lv_Komsa_2023}. We refer to this method as \gls{dse}.

\subsection{Wick's theorem}
$G_{\mathbf{q};\mathbf{k}}^{\nu\nu';\mu\mu'}(t)$ can be decomposed using Wick's approximation \cite{Bloch_De_Dominicis_1958},
\begin{equation}\label{eq:wicks}
    \begin{split}     
        &\braket{A_{\mathbf{q}\nu}(t)A_{-\mathbf{q}\mu}(t)A_{\mathbf{k}\nu'}^{\dagger}(0)A_{-\mathbf{k}\mu'}^{\dagger}(0)}\\
         &\approx\braket{A_{\mathbf{q}\nu}(t)A_{-\mathbf{q}\mu}(t)}\braket{A_{\mathbf{k}\nu'}^{\dagger}(0)A_{-\mathbf{k}\mu'}^{\dagger}(0)}\\
         &+ \braket{A_{\mathbf{q}\nu}(t)A_{\mathbf{k}\nu'}^{\dagger}(0)}\braket{A_{-\mathbf{q}\mu}(t)A_{-\mathbf{k}\mu'}^{\dagger}(0)} \\
         &+ \braket{A_{\mathbf{q}\nu}(t)A_{-\mathbf{k}\mu'}^{\dagger}(0)}\braket{A_{-\mathbf{q}\mu}(t)A_{\mathbf{k}\nu'}^{\dagger}(0)}.
    \end{split}
\end{equation}
This allows us to express higher order scattering in terms of convolutions of second order scattering.
Notably, the same theorem can be applied to all even orders.

\section{Modeling details} \label{sec:modeling-details}

\subsection{Potential energy surface}

A machine-learned \gls{pes} for \gls{bzo} was developed in Ref. \cite{FraRosErhWah2023} using the \gls{nep} approach. It was trained from \gls{dft} data using the van der Waals density functional with consistent exchange (vdW-DF-cx) \cite{Dion_Rydberg_Schröder_Langreth_Lundqvist_2004,Berland_Hyldgaard_2014} for the exchange-correlation effects, here denoted CX. This functional gives a good balance between accuracy and computational speed. However, it consistently underestimates the vibrational frequencies at the $\Gamma$-point with about 5\% \cite{FraRosErhWah2023}. For more details and comparisons with other functionals, see Refs \cite{Perrichon_Jedvik_Granhed_Romanelli_Piovano_Lindman_Hyldgaard_Wahnström_Karlsson_2020,FraRosErhWah2023}. CX was used in Ref. \cite{FraRosErhWah2023} to study the structure and dynamics of \gls{bzo} as function of temperature and pressure.
Additionally, a phase transition from the cubic to the tetragonal structure was obtained at \qty{16.2}{\giga\pascal} at ambient temperature.

\subsection{Dielectric susceptibility}

To obtain the Raman spectrum in \autoref{eq:lineshape} a model for the dielectric susceptibility is also required. 
That was developed in Ref. \cite{Xu_Rosander_2023} by generalizing the NEP scheme to enable predictions of tensorial properties, the \gls{tnep} approach. 
It was based on \gls{dft} data for the relative susceptibility using the CX functional.
The training structures were generated by running \gls{md} with different temperatures and pressures with the same \gls{nep} model as used in Ref. \cite{FraRosErhWah2023}. 
Various sizes of the supercell were used, with the total number of atoms ranging from 5 to 40 atoms.
For more details, see Ref. \cite{Xu_Rosander_2023}.

\subsection{Molecular dynamics}

In the present study, the \gls{md} simulations are done using the \gpumd{} package \cite{Fan_Chen_Vierimaa_Harju_2017} together with the NEP in Ref.\cite{FraRosErhWah2023} and the \gls{tnep} in Ref. \cite{Xu_Rosander_2023} to obtain the correlation function in \autoref{eq:lineshape}. 
In all simulations, we employ a time step of \qty{1}{\femto\second}. 
The system is equilibrated during a period of \qty{100}{\pico\second} in the NVT ensemble. 
The time correlation functions are then sampled over \qty{500}{\pico\second} in the NVE ensemble, and averaged over 20 identical simulations. 
Lattice parameters are obtained from NPT simulations, as done in Ref.~\cite{FraRosErhWah2023}. 
For the Raman simulation at ambient pressure (\autoref{fig:raman_simulated_experiment}) we have used a $14\times14\times14$ supercell (13 720 atoms), while for the pressure dependent calculations (\autoref{fig:raman_pressure}, \ref{fig:raman_pressure_wide_range} and \ref{fig:raman_pressure_range_zoom}) a $12\times12\times12$ supercell is used (8 640 atoms).
For the R-mode calculations (\autoref{fig:modes_in_time}, \ref{fig:Power_spectrum} and \ref{fig:frequencyVpressure}) we used a $24\times24\times24$ supercell (69 120 atoms).

%
\end{document}